\documentclass[preprint,authoryear,12pt,]{elsarticle}
\usepackage{epsfig,graphicx,amssymb,amsmath,times,natbib}

\journal{Advances in Space Research}
\begin{document}
\begin{frontmatter}
\title{Multi-wavelength study of the short term TeV flaring activity from the blazar Mrk 501 observed in June 2014}
\author[1]{K K Singh\corref{cor}}
\cortext[cor]{Corresponding author}
\ead{kksastro@barc.gov.in}
\author[1]{H Bhatt} \author[1,2]{S Bhattacharyya} \author[1]{N Bhatt} \author[1,2]{A K Tickoo} \author[1,2]{R C Rannot}
\address[1]{Astrophysical  Sciences  Division, Bhabha Atomic Research Centre, Mumbai - 400 085, India}
\address[2]{Homi Bhabha National Institute, Mumbai - 400 094, India}
\begin{abstract}
In this work, we study the short term flaring activity from the high synchrotron peaked blazar Mrk 501 
detected by the {\bf FACT} and H.E.S.S. telescopes in the energy range 2-20 TeV during June 23-24, 2014 (MJD 56831.86-56831.94).
We revisit this major TeV flare of the source in the context of near simultaneous multi-wavelength observations 
of $\gamma$--rays in MeV-GeV regime with \emph{Fermi}-LAT, soft X-rays in 0.3-10 keV range with 
\emph{Swift}-XRT, hard X-rays in 10-20 keV and 15-50 keV bands with MAXI and \emph{Swift}-BAT respectively, 
UV-Optical with \emph{Swift}-UVOT and 15 GHz radio with OVRO telescope. We have performed a detailed temporal 
and spectral analysis of the data from \emph{Fermi}-LAT, \emph{Swift}-XRT and \emph{Swift}-UVOT during the period
June 15-30, 2014 (MJD 56823-56838). Near simultaneous archival data available from \emph{Swift}-BAT, MAXI and 
OVRO telescope along with the V-band optical polarization measurements from SPOL observatory are also used in the 
study of giant TeV flare of Mrk 501 detected by the {\bf FACT} and H.E.S.S. telescopes. No significant change in the 
multi-wavelength emission from radio to high energy $\gamma$--rays during the TeV flaring activity of Mrk 501 is 
observed except variation in soft X-rays. The varying soft X-ray emission is found to be correlated with the 
$\gamma$--ray emission at TeV energies during the flaring activity of the source. The soft X-ray photon spectral index is 
observed to be anti-correlated with the integral flux {\bf showing harder-when-brighter behavior}. An average value of 
4.5$\%$ for V-band optical polarization is obtained during the above period whereas the corresponding electric vector 
position angle changes significantly.{\bf We have used the minimum variability timescale from the H.E.S.S. observations to 
estimate the Doppler factor of the emission region which is found to be consistent} with the previous studies of the source.        
\end{abstract}

\end{frontmatter}
\section{Introduction}
Blazars are a special class of radio loud active galactic nuclei (AGN) with powerful and highly relativistic jets 
oriented at small angles ($\le10^\circ$) to the line of sight of the observer on the Earth \citep{Readhead1978, Urry1995}. 
The jets of blazars are assumed to be powered by rotating supermassive black holes surrounded by accretion disks 
at the centers of massive elliptical type host galaxies. They are observed to be highly luminous at all wavelengths 
throughout the electromagnetic spectrum from radio to very high energy (VHE, E$>$100 GeV) $\gamma$--rays. The multi-wavelength 
emission from blazars is dominated by relativistic effects like Doppler boosting of the observed flux and dilation/compression 
of timescales due to orientation of the jets. The observed broadband spectral energy distribution (SED) of blazars from radio 
to $\gamma$--rays is characterized by two broad peaks. The origin of non-thermal SED is attributed to the relativistic 
charged particles in the blazar jets. The physical process responsible for the low energy peak at UV/optical to soft X-ray is 
relatively well understood and assumed to be the synchrotron radiation of relativistic electrons in the tangled magnetic field 
of the jet \citep{Urry1982, Marscher2008}. A significant contribution to the low energy hump has also been observed in the SED of 
many blazars in the narrow energy range from non-jet components like accretion disk, broad line region (BLR) and dusty torus 
\citep{Kushwaha2014, Nalewajko2014}. 
\par
The origin of high energy (HE, $>$ 30 MeV) component of the SED peaking at hard X-rays to MeV-GeV $\gamma$--rays is not 
well understood and two alternative approaches: \emph{leptonic} and \emph{hadronic} have been proposed. In the leptonic 
approach, the origin of HE component of the SED is attributed to the inverse Comptonization of different circumnuclear 
low energy seed photons produced within the jet or outside it by relativistic leptons. If seed photons for the 
inverse Compton upscattering are the synchrotron photons produced by the same population of relativistic leptons 
within the jet, the process is referred to as synchrotron self Compton (SSC) model \citep{Maraschi1992, Sokolov2004}. 
On the other hand, if the seed photon field is external to the jet, namely from accretion disk, torus, broad line region 
or cosmic microwave background radiation, the process is termed as external Comptonization (EC) model \citep{Dermer1992, 
Sikora1994, Agudo2011}. Alternatively, in the hadronic approach, the emission of high energy component of the blazar SED 
is attributed to proton synchrotron or secondary emission from p-$\gamma$ interactions \citep{Mannheim1993, Pohl2000, Aharonian2002}. 
However, the leptonic models are more favored than hadronic processes for modelling the rapid variability observed in $\gamma$--ray 
emission from blazars \citep{Bottcher2013, Cerruti2015}. 

\par
Blazars are classified into two broad categories: BL Lacertae objects (BL Lacs) and Flat Spectrum Radio Quasars (FSRQs) 
on the basis of the strength of emission lines in their optical spectra. The optical spectra of BL Lacs are characterized 
as featureless continuum emission with very weak or no emission lines, whereas FSRQs have prominent and broad emission 
lines \citep{Urry1995, Stocke1991, Marcha1996}. The position of peak frequency in the low energy component of the SED 
has also been used to subdivide blazars in three subclasses namely high synchrotron peaked (HSP), intermediate 
synchrotron peaked (ISP) and low synchrotron peaked (LSP) blazars \citep{Abdo2010}. The location of synchrotron peak 
frequency for HSPs is observed at UV/X-ray energies whereas FSRQs and ISPs have peak at IR/optical energies. The peak of HE 
component of the SED lies at GeV-TeV energies for HSPs and at hard X-ray energies for FSRQs and LSPs. BL Lacs and FSRQs 
have low and high luminosity respectively-forming the so called blazar sequence, but there is no general consensus on 
this phenomenological feature of blazars \citep{Ghisellini2017}. 
     
\par 
Emissions from most of the blazars have been observed to be variable over the entire electromagnetic spectrum at 
short timescales during the flaring activities followed by quiescent state. The multi-wavelength emissions from 
blazars show a high degree of polarization and strong variability on timescales from days to minutes. Rapid flaring 
activities in blazars on hour and minute timescales are dominant at $\gamma$--ray energies with change in flux up to 
few magnitudes \citep{Falomo2014, Ackermann2016, Rani2013}. In general, the flaring activity in X-ray and VHE regimes 
has become one of the most important observational features of the blazars. However, the physical processes involved 
in the origin of such dramatic behaviour of blazars have not been clearly explained. In this paper, we present the 
multi-wavelength study of the short term TeV flaring activity from the blazar Mrk 501 detected by the H.E.S.S. and FACT telescopes 
\citep{Chakraborty2015, Cologna2016} during the night of June 23-24, 2014 (MJD 56831.86-56831.94) . 
In Section 2, we briefly describe the observational history of the blazar Mrk 501 over the last two decades. 
The details of multi-wavelength observations and data analysis are given in Section 3. Section 4 provides 
discussion of the results obtained in this study. Finally, we summarize the important findings of this work in Section 5. 
We have assumed  a flat $\Lambda$CDM cosmology with $\Omega_m = 0.3$, $\Omega_\Lambda = 0.7$ and 
$H_0 = 70\;km\;s^{-1}\;Mpc^{-1}$ throughout the paper.        
  
\section{Mrk 501}
Mrk 501 is one of the nearest (z=0.034, 145 Mpc) and the brightest HSP blazars at X-ray and TeV energies in 
the extragalactic Universe. This source has been observed over the last two decades to undergo major outbursts 
on long timescales and rapid short timescales prominently in X-ray and TeV energy bands. Therefore, most of the 
$\gamma$--ray studies of Mrk 501 in the past have been performed during its flaring activity. Both low and high 
energy peaks of Mrk 501 SED are observed to shift towards higher energies during different flaring activities 
of the source. The first TeV emission from Mrk 501 was detected in 1995 by the Whipple telescope above 300 GeV 
\citep{Quinn1996} and subsequently confirmed by the HEGRA group \citep{Bradbury1997}. In 1997, it was observed in 
high state of activity by the CAT observatory at energies above 250 GeV with flux level two orders of magnitude 
higher than its discovery level \citep{Djannati1999}. During the same time, the source was also observed in flaring 
state at soft X-ray energies by All Sky Monitor (ASM). In 2005, the MAGIC telescope observed Mrk 501 in flaring state 
with fastest VHE $\gamma$--ray variability on a timescale of minutes \citep{Albert2007}. Between March and May 2008, 
a multi-wavelength campaign was planned on Mrk 501 to perform an unbiased characterization of $\gamma$--ray emission 
together with X-ray to radio emission from the source in the low state of activity \citep{Aleksic2015}. The VHE flux 
during this campaign was found to be 10$\%$--20$\%$ of the flux level measured from the flaring activity of Mrk 501. 
\par
In 2009, \emph{Fermi}-LAT observed remarkable spectral variability and mild flux variability in the energy range 
100 MeV--300 GeV from Mrk 501 during first 480 days of \emph{Fermi} operation \citep{Abdo2011}. In March 2009, 
a short-term multi-wavelength observation of Mrk 501 was coordinated to provide a baseline measurement of the source in 
the quiescent state including the TeV observations with VERITAS and MAGIC \citep{Acciari2011}. A major VHE outburst was 
observed on May 1, 2009 by Whipple and VERITAS during an extensive multi-wavelength campaign of Mrk 501 between 
April 17 and May 5, 2009 \citep{Aliu2016}. In the flaring state, the VHE flux above 400 GeV increased to five 
times the Crab Nebula flux and ten times the pre-flare flux showing a fast flux variation with an increase of a factor 4 
in 25 minutes and decay time of 50 minutes \citep{Aliu2016}. In 2011, the brightest TeV flare from Mrk 501 since 2005 was 
observed by ARGO-YBJ experiment \citep{Bartoli2012}. An increase of $\gamma$--ray flux above 1 TeV by a factor of $\sim$ 7 
from its quiescent state emission had been detected by ARGO-YBJ during this period. Again in May 2012, Mrk 501 was 
observed in relatively high $\gamma$--ray emission state by TACTIC telescope in the energy range above 850 GeV \citep{Chandra2017}. 
\par
First detailed study of the synchrotron peak in the SED of Mrk 501 was reported using broad band observations of the source 
during April-August 2013 \citep{Furniss2015}. In this study, the hard X-ray observations with NuSTAR provided unprecedented evidence 
for the rapid variability at hour timescales in the energy range 3-79 keV. During March-October 2014, short term strong X-ray flares 
with flux level varying by factors of 2-5 were detected from Mrk 501 on timescales of few weeks or shorter \citep{Kapanadze2017}. 
The X-ray flux points in the energy range 0.3-10 keV were observed to be correlated with the TeV flux measurements during the 
flaring activity in the source. Apart from the TeV-X-ray correlation, more complicated variability patterns had also been observed 
which indicated that $\gamma$--ray emission region in Mrk 501 is more complex than a single blob \citep{Kapanadze2017}. 
Different features at TeV energies in low and high activity states of Mrk 501 have been reported by various observational studies 
but the physical mechanism involved is not yet completely understood.     

\section{Multi-Wavelength observations and Data Analysis}
In 2014, the H.E.S.S. telescopes observed Mrk 501 on several nights between June 19-25 and on July 29-30. A major flaring activity 
from the source was detected during the night of June 23-24, 2014 (MJD 56831.86-56831.94) with the highest flux level of Mrk 501 ever 
observed by H.E.S.S. \citep{Chakraborty2015}. The model analysis of H.E.S.S. data yielded an excess of $\sim$ 1200 $\gamma$--ray photons 
with statistical significance of 67$\sigma$ for 2.1 hours of observations on the night of June 23-24, 2014. This unprecedented flaring 
activity from Mrk 501 with variability timescale of $\sim$ 6 minutes \citep{Chakraborty2015} in the energy range 2-20 TeV provides a unique 
opportunity to investigate the emission mechanism from high synchrotron peaked blazars like Mrk 501. Near simultaneous observations of the 
flaring activity of Mrk 501 with FACT \citep{Cologna2016} above 750 GeV provided important pre- and post-flare information with dense sampling. 
In the night of June 23-24, 2014 (MJD 56831.86-56831.94), FACT observed Mrk 501 for a live time of $\sim$ 4 hours with flux level more than four 
times the Crab Nebula flux above 750 GeV allowing for better characterization of the flaring activity in the VHE band. A simple linear correlation 
was observed between the nightly averaged flux points including the flaring activity recorded with H.E.S.S. and FACT telescopes \citep{Cologna2016}.   
With the motivation of understanding the broadband emission of Mrk 501 during the short term giant flaring activity, we use near simultaneous 
multi-wavelength data in the lower energy bands from HE $\gamma$--ray to radio observations of the source during the period June 15-30, 2014 
(MJD 56823-56838). A brief description of the details of multi-wavelength data from various instruments world wide and their analysis procedure is given below.

\subsection{\emph{Fermi}-LAT}
The Large Area Telescope (LAT) on board \emph{Fermi} satellite is a pair-conversion $\gamma$--ray telescope with wide field of view of 
2.4 steradians and can detect photons of energy above 100 MeV with an energy resolution of 10$\%$ at 1 GeV in survey mode \citep{Atwood2009}. 
We have downloaded the Pass 8 \emph{Fermi}-LAT data for Mrk 501 from \emph{Fermi} Science Support Center 
(FSSC)\footnote{https://fermi.gsfc.nasa.gov/cgi-bin/ssc/LAT/LATDataQuery.cgi} during the period June 15-30, 2014 including the H.E.S.S. observations 
of the source in flaring state. The LAT data have been analysed using publicly available \emph{Fermi} ScienceTools v10r0p5. We have selected photons 
from a circular region of interest of 15$^\circ$ radius centered on the position of Mrk 501 using Pass 8 response function (event class=128, event type=3) 
corresponding to $P8R2\_SOURCE\_V6$ response in the energy range 100 MeV--300 GeV. In addition, we use maximum zenith angle cut of 90$^\circ$ to avoid the 
contamination from Earth limb $\gamma$--rays and spacecraft rocking angle $>$ 52$^\circ$ to limit the time periods when Earth entered LAT field of view. 
We have constructed the sky source model file using third \emph{Fermi} $\gamma$--ray catalog (3FGL) \citep{Acero2015}. The Galactic diffuse and 
extragalactic isotropic background emissions have been modelled using $gll\_iem\_v06.fits$ and $iso\_P8R2\_SOURCE\_v6\_v06.txt$ files respectively. 
The spectral parameters of the source and detection significance for daily observations are estimated using an unbinned maximum-likelihood implemented 
in \emph{gtlike} tool. The daily fluxes for the light curve of the source have been estimated using power law model keeping spectral index and normalization 
as free parameters. For daily observations with test static (TS) less than 10, we have calculated 2$\sigma$ upper limit on the integral flux above 100 MeV.  
\par
We have also performed the spectral analysis of Mrk 501 in the energy band 100 MeV--300 GeV for the time interval June 22-24, 2014 
(MJD 56830-56832) overlaping with flare period observed with the H.E.S.S. telescopes. The spectrum of the source for this time period is determined by 
dividing the energy band 100 MeV--300 GeV into three smaller energy bins : 0.1--1 GeV, 1--5 GeV and 5--300 GeV. A power law model is fit 
to the data in three energy bins with the normalization as free parameter and spectral index set to the value obtained from the 
integration of data for the above time interval in the energy range 100 MeV--300 GeV. 

\subsection{\emph{Swift}-BAT}
The Burst Alert Telescope (BAT) on board \emph{Neil Gehrels Swift Observatory} provides an all-sky hard X-ray survey covering the energy 
range 15-150 keV \citep{Barthelmy2005}. The instrument is a highly sensitive, coded mask imager with large field of view and provides 
positional accuracy between 1-3 arcmin. The \emph{Swift}-BAT hard X-ray transient monitor \citep{Krimm2013} provides X-ray observations in 
the energy range 15-50 keV and generates light curves of sources spanning over more than thirteen years. We have downloaded daily light curve 
of Mrk 501 for the period June 15-30, 2014 from online \emph{Swift}-BAT  transient monitor program\footnote{https://swift.gsfc.nasa.gov/results/transients}.

\subsection{MAXI}
The Monitor of All-sky X-ray Image (MAXI) is a payload on Japanese Experimental Module on the International Space Station 
to investigate the long term behaviour of X-ray sources in the energy range 2-20 keV \citep{Matsuoka2009}. This mission has 
two X-ray detectors consisting of gas proportional counters and X-ray charge coupled devices with the ability to make an all-sky X-ray 
map at soft and medium energy X-rays. The instrument has poor angular resolution, but can acheive localization accuracy up to 
0.1$^\circ$ for bright sources. We have obtained daily observations of Mrk 501 with MAXI in the energy range 10-20 keV during 
the period  June 15-30, 2014 from the online data archive provided by MAXI team\footnote{http://maxi.riken.jp/pubdata/}.   

\subsection{\emph{Swift}-XRT}
The X-Ray Telescope (XRT) on board the \emph{Swift} satellite is a grazing incidence Wolter I telescope with CCD detector 
covering the energy range 0.3--10 keV with effective area of 110 cm$^{2}$ and field of view of $\sim$ 23.6 arcmin \citep{Burrows2005}.
The archival data  on Mrk 501 have been downloaded from  the \emph{Swift} multi-wavelength support 
program\footnote{http://www.swift.psu.edu/monitoring/} for the observation period June 15-30, 2014. We have analyzed 
the \emph{Swift}-XRT data to derive the spectral features of the source during these periods using \emph{HEASoft} package 
version 6.19. The observations have been performed in windowed timing (WT) mode and the cleaned XRT event files with 
standard filtering criteria are produced using {\sc xrtpipeline} version 0.13.2 with recent calibration files (version 20160609). 
The spectra of the source and the background are generated from {\sc Xselect V2.4d} in the energy band 0.3-10 keV. The source 
photons have been extracted from a circular region with radius $\rm{45^{\prime\prime}}$ around source position. The nearby 
background with similar radius from the source free region is selected for individual observation according to the source 
position in the detector. The ancillary response files (ARFs) have been generated with the {\sc xrtmkarf} task, applying 
corrections for the point spread function losses and CCD defects using the cumulative exposure map. Finally, the observations 
are binned to have atleast 30 counts per spectral bin using {\sc grppha}. 
\par
The spectrum has been fitted using power law model with absorption due to a neutral hydrogen using {\sc xspec} (ver 12.9) 
model {\sc phabs}$\times${\sc zpow} for source redshift $z=0.034$ in the energy range 0.3-10 keV. The line-of-sight absorption 
is fixed to a neutral hydrogen column density ($N_H$) of $\rm{1.55\times10^{20}~cm^{-2}}$ obtained from the Leiden-Argentine-Bonn 
(LAB) survey of Galactic HI \citep{Kalberla2005}. The analysis results for each observations during the period June 15-30, 2014 
(MJD 56823-56838) have been summarized in Table \ref{tab:xrt}. We have also estimated the energy flux values in four sub-energy 
bands: 0.3-0.7 keV, 0.7-1.7 keV, 1.7-4.0 keV and 4.0- 10.0 keV for each observation to derive the spectral energy distribution 
of the source and to study the spectral evolution during this period. We have also tried to fit the spectrum using log-parabola 
model instead of the simple power law because some of the $\chi^2$-values reported in \ref{tab:xrt} are high for power law fit. 
The fit improves marginally with almost no change in the photon spectral index parameter ($\alpha$) but the curvature index ($\beta$) 
has large uncertainty. Therefore, we have used simple power law model to fit the spectrum in the energy range 0.3-10 keV in the present study.   
\begin{table}
\caption{\emph{Swift}-XRT spectral analysis of the data during the period June 15-30, 2014 using power law model ($F \propto E^{-\alpha}$) 
		with absorption and neutral hydrogen density fixed at $1.55\times10^{20}~cm^{-2}$.}
\label{tab:xrt}
\vspace{0.3cm}
\begin{center}
\begin{tabular}{lcccccc}
\hline
Obs ID  	&MJD start	&Exposure	&Photon index			&$\chi^2$/dof\\ 
		&		&(sec)		&($\alpha$)\\	
\hline
00030793249     &56824.729	&1470		&1.76$\pm$0.01			&305/274 \\
00035023038     &56826.191	&963		&1.70$\pm$0.01			&269/240 \\ 
00035023039     &56829.000   	&722		&1.58$\pm$0.02			&229/233 \\ 
00030793253     &56830.004      &681	        &1.72$\pm$0.02			&211/186 \\ 
00030793252     &56830.998      &683	 	&1.67$\pm$0.02			&221/216 \\ 
00035023040     &56832.004      &1137	        &1.60$\pm$0.01			&274/265 \\ 
00035023041     &56832.253      &620		&1.64$\pm$0.02			&221/199 \\
00035023042     &56832.982      &315	        &1.73$\pm$0.03			&99/78  \\
00035023043     &56834.050      &833		&1.69$\pm$0.02			&234/219 \\
00035023044  	&56834.982  	&1246		&1.73$\pm$0.01			&371/262 \\
00035023045     &56835.249   	&958	        &1.69$\pm$0.01			&280/232 \\
00035023047  	&56836.982  	&1519		&1.88$\pm$0.01			&350/256 \\
\hline
\end{tabular}
\end{center}
\end{table} 

\subsection{\emph{Swift}-UVOT}
The Ultraviolet/Optical Telescope (UVOT) on board the \emph{Swift} observatory  is a modified optical configuration having 
micro-channel plate intensified CCD detectors to provide observations in the wavelength range 170-600 nm \citep{Roming2005}. 
\emph{Swift}-UVOT utilizes three optical filters V(546.8 nm), B(439.2 nm), U(346.5 nm) and three ultraviolet filters 
UVW1(260.0 nm), UVM2(224.6 nm), UVW2(192.8 nm) during the observations. We have downloaded the online available 
raw data files from the \emph{Swift}-UVOT observations of Mrk 501 during the period June 15-30, 2014 from 
 the archive\footnote{http://www.swift.psu.edu/monitoring/}. The data available in each filter has been analyzed 
using {\sc uvotsource} task. A circular region of 13$^{\prime\prime}$ radius is selected around the source to 
extract the UVOT source counts, while extracting the background for two circular regions of radius 30$^{\prime\prime}$ each 
from the source free regions near to the source. The approximate aperture corrections have been applied by specifying 
``apercorr=CURVEOFGROWTH'' when running the task {\sc uvotsource}. Finally, the magnitudes of the source measured by 
UVOT in the six filters have been converted to energy fluxes for each observation available during the above period 
using the corresponding zero flux points given in \citep{Poole2008}. 
      
\subsection{SPOL}
The optical spectropolarimetric (SPOL) monitoring program at Steward observatory \citep{Smith2009} has been designed to measure the 
linear polarization and flux from $\gamma$--ray bright blazars and to provide the data to the research community 
under \emph{Fermi} Multiwavelength Observing-Support Programs\footnote{https://fermi.gsfc.nasa.gov/ssc/observations/
multi/programs.html}. SPOL is dual-beam spectropolarimeter with a waveplate and Wollaston prism to modulate and analyze 
polarized light and provides spectral coverage in the wavelength range 400--755 nm with a resolution of about 2 nm \citep{Smith2009}.
We have obtained the polarization data for Mrk 501 during the period June 15-30, 2014, publicly available from SPOL 
observations of bright \emph{Fermi} blazars\footnote{http://james.as.arizona.edu/psmith/Fermi}.

\subsection{OVRO}
The Owens Valley Radio Observatory (OVRO) 40 m monitoring program \citep{Schmidt1992} provides blazar observations at 15 GHz under 
\emph{Fermi} Multiwavelength Observing-Support Programs. The OVRO 40 m telescope is a f/0.4 parabolic reflector with its receiver 
operating in Ku band at central frequency of 15 GHz, bandwidth of 3 GHz and a noise-equivalent reception bandwidth of 2.5 GHz \citep{Richards2011}. 
We have downloaded publicly available data from OVRO archive\footnote{http://www.astro.caltech.edu/ovroblazars/data} for Mrk 501 observations 
during the period June 15-30, 2014.
\section{Results and Discussion}

\subsection{Multi-wavelength light curve}
\begin{figure}
\begin{center}
\includegraphics[width=1.0\textwidth]{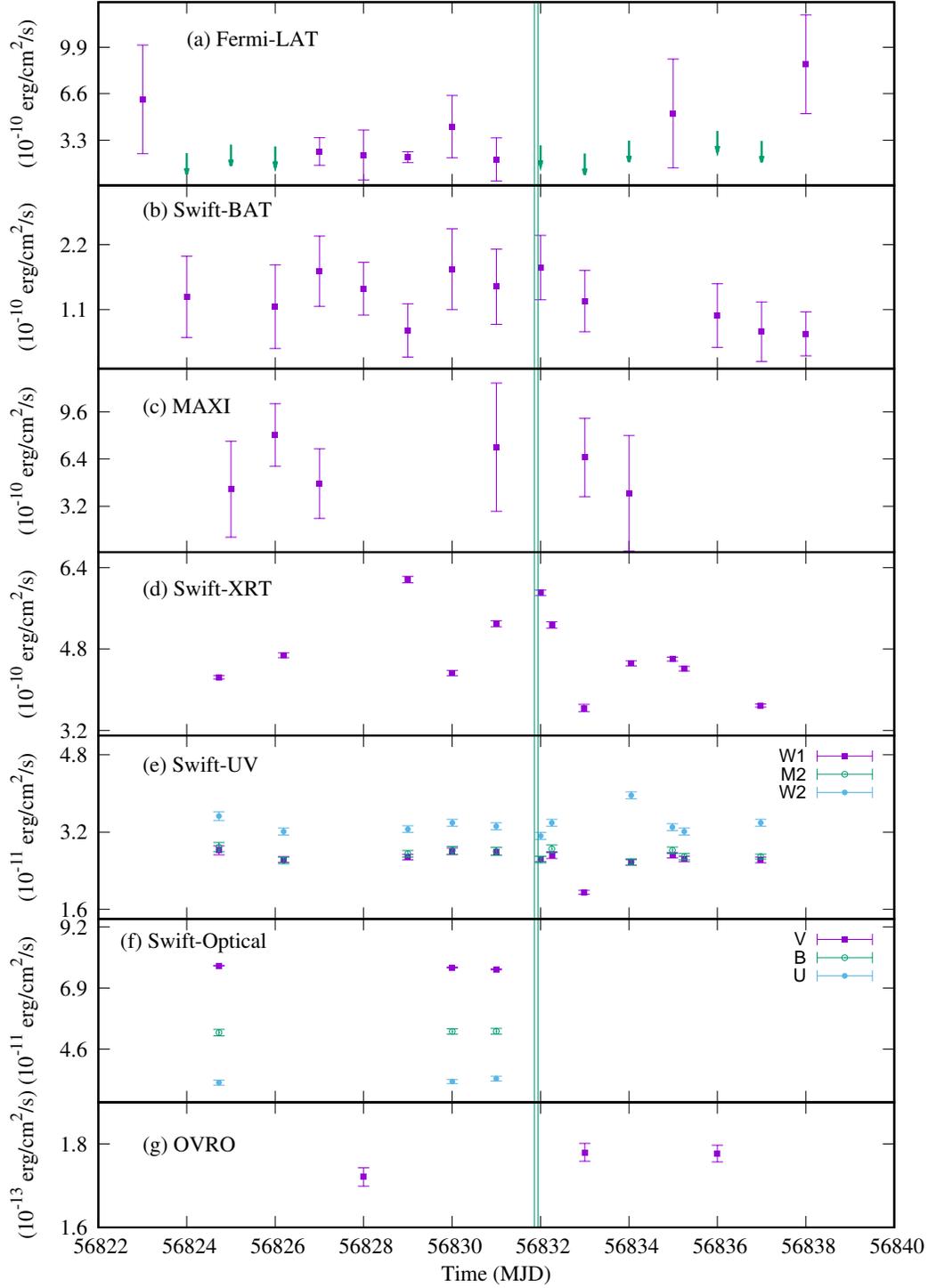}
\caption{Daily averaged multi-wavelength light curves of Mrk 501 during the period June 15-30, 2014 (MJD 56823-56838) 
	observed with different instruments from HE $\gamma$--rays to radio. The error bars in XRT, UV and optical 
        flux measurements are very small and their size is comparable to the size of plot points. The time interval 
        between two vertical lines represents the period ($\sim$ 2.1 hours) of  TeV flaring activity  of the source 
        detected by H.E.S.S. on the night of June 23-24, 2014 (MJD 56831.86-56831.94) \citep{Chakraborty2015}.}
\label{fig:lc}
\end{center}
\end{figure}

The multi-wavelength light curves of Mrk 501 from HE $\gamma$--rays to radio for the period June 15-30, 2014 (MJD 56823-56838) covering the period 
of VHE $\gamma$--ray flare on the night of June 23-24, 2014 (MJD 56831.86-56831.94) are shown in Figure \ref{fig:lc}(a-g). The duration of short term 
TeV flaring activity detected by the H.E.S.S. telescopes is indicated by two vertical lines in Figure \ref{fig:lc}. The values of average emission in 
all the energy bands estimated from the constant fit to the light curves and corresponding reduced $\chi^2_r$ (dof: degree of freedom) are given in Table 
\ref{tab:constant}. We observe that the multi-wavelength emission of the blazar Mrk 501 from HE $\gamma$--rays to radio is consistent with constant emission 
except for soft X-ray emission detected by \emph{Swift}-XRT during this period. The daily averaged soft X-ray flux in the energy range 0.3-10 keV measured by 
\emph{Swift}-XRT is shown in Figure \ref{fig:lc}(d). It is evident that near simultaneous X-ray activity of the source in the energy range 0.3-10 keV is 
consistent with the VHE activity in the energy range 2-20 TeV detected with H.E.S.S. (Figure 1 in \citet{Chakraborty2015}) and above 750 GeV observed 
with FACT (Figure 1 in \citet{Cologna2016}). In soft X-ray band, the source is observed in the highest activity state on June 21, 2014 (MJD 56829), but no 
contemporeneous activity is detected in VHE band by the FACT and H.E.S.S. telescopes. In VHE band, the source is observed in extreme activity state on the night of 
June 23-24 (MJD 56831.86-56831.94) with near simultaneous high state emission in soft X-rays. The variability of the source in VHE band is observed to be very fast 
with minimum variability timescale of $\sim$ 6 minutes \citep{Chakraborty2015} in the observations of flaring activity with H.E.S.S. but no minute scale variability 
is obtained in the FACT observations \citet{Cologna2016} due to low sensitivity of the telescope. With this rapid variability of the source, the absence of VHE flare 
during the highest soft X-ray activity on June 21, 2014 (MJD 56829) can be attributed to the non-simultaneous observations with H.E.S.S., FACT and \emph{Swift}-XRT. 
However, X-ray emissions in the energy bands 10-20 keV (Figure \ref{fig:lc}(c)) and 15-50 keV (Figure \ref{fig:lc}(b)) detected with MAXI and \emph{Swift}-BAT 
respectively are consistent with constant emission within statistical uncertainties. 
\par
The HE $\gamma$--ray emission of Mrk 501 observed with \emph{Fermi}-LAT as shown in Figure \ref{fig:lc}(a), also does not show any signature of variability 
during this period. The flux points with downward arrow in the Figure \ref{fig:lc}(a) represent 2$\sigma$ upper limits on the integral flux above 100 MeV for 
the observations when source is not significantly detected with \emph{Fermi}-LAT. The constant emissions in UV, optical and radio bands depicted in Figure 
\ref{fig:lc}(e-g) respectively during the period June 15-30, 2014 (MJD 56823-56838) are consistent with the source behaviour known from the past observations. 
A positive correlation between VHE $\gamma$--ray emission above 750 GeV observed with FACT and X-rays in the energy range 2-10 keV had been reported by 
\citet{Cologna2016} for the above observation period. But, the correlation was violated during the TeV flaring activity observed on the night of 
June 23-24, 2014 \citep{Cologna2016, Cologna2015}. This suggests that the physical parameters involved in the VHE $\gamma$--ray emission during the extreme 
flaring state may be different from the rest of the emissions observed from Mrk 501 \citet{Cologna2016}. Different values of correlation index between 
the TeV and X-ray emissions can be derived depending on the assumed evolution scenario in different spectral bands under the frame-work of homogeneous 
single zone SSC model \citep{Katarzynski2005}. A quadratic or more than quadratic correlation between the TeV and X-ray flux points is expected for 
simultaneous emission from two independent zones, where one zone emits in X-ray band and other zone dominates at TeV $\gamma$--rays \citep{Katarzynski2010}.
However, a quadratic correlation between soft X-ray and TeV $\gamma$--ray flux points during the flaring activity can be explained for a specific choice under 
the framework of single zone SSC model when the source activity is attributed to the variation in the leptonic particle density \citep{Singh2017, Sahayanathan2018}. 
In this case, TeV flux due to SSC process varies quadratically with respect to X-ray flux produced by the synchrotron process. Referring to the VHE flare detected 
by the H.E.S.S. telescopes, the TeV light curve in the energy range 2-20 TeV has been divided into two sub energy bands: 2-4.5 TeV and 4.5-20 TeV 
(Figure 2 in \citet{Chakraborty2015}). Variations at short timescale are observed in both energy bands. The detailed temporal analysis of the VHE light curve is beyond 
the scope of the present study. 
  
\begin{table}
\caption{Constant emission model fit to the multi-wavelength light curves of Mrk 501 during the period June 15-30, 2014 (MJD 56823-56838) 
         as shown in Figure \ref{fig:lc}.}
\label{tab:constant}
\vspace{0.3cm}
\begin{center}
\begin{tabular}{lcccc}
\hline
Energy band  	&Instrument	&Constant flux (erg~$cm^{-2}$~s$^{}$) 	&$\chi^2_r$/dof\\ 
\hline
0.1-300 GeV	&Fermi/LAT	&(2.30$\pm$0.31)$\times10^{-10}$	&0.86/7\\
15-50  keV	&Swift/BAT	&(1.17$\pm$0.12)$\times10^{-10}$	&0.66/11\\
10-20  keV	&MAXI		&(6.10$\pm$0.68)$\times10^{-10}$	&0.36/5\\
0.3-10 keV	&Swift/XRT	&(4.52$\pm$0.26)$\times10^{-10}$	&234/11\\
203 nm (W2 band)&Swift/UVOT	&(3.36$\pm$0.07)$\times10^{-11}$	&9.8/10\\
223 nm (M2 band)&Swift/UVOT	&(2.72$\pm$0.03)$\times10^{-11}$	&2/10\\
263 nm (W1 band)&Swift/UVOT	&(2.54$\pm$0.09)$\times10^{-11}$	&25/11\\
350 nm (U band)	&Swift/UVOT	&(3.40$\pm$0.05)$\times10^{-11}$	&0.78/2\\
432 nm (B band) &Swift/UVOT	&(5.25$\pm$0.02)$\times10^{-11}$	&0.49/2\\
540 nm (V band)	&Swift/UVOT	&(7.68$\pm$0.04)$\times10^{-11}$	&0.13/2\\
15 GHz (Radio)	&OVRO		&(1.76$\pm$0.02)$\times10^{-13}$	&2.34/2\\		
\hline
\end{tabular}
\end{center}
\end{table}

\subsection{Variability analysis}

The variability present in the flux values simultaneously measured in different energy bands would provide 
useful information regarding the physical processes involved in the emission mechanism at the source. In order 
to search for the intrinsic variability in the near simultaneous multi-wavelength light curves obtained for the 
period June 15-30, 2014 during the TeV flaring activity of Mrk 501, we have estimated the fractional variability 
amplitude (F$_{var}$) parameter. We have used the description proposed in \citep{Vaughan2003} to estimate the fractional 
variability amplitude in different energy bands from radio to HE $\gamma$--rays. F$_{var}$ is defined as 
\begin{equation}
	F_{var}=\frac{\sqrt{S^{2}-\langle \sigma_{err}^{2}\rangle}}{\langle F\rangle} 
\end{equation}
where $\langle F\rangle$ is the average flux, \textit{S} is the standard deviation and $\sigma_{err}^{2}$ is  
the mean square error of N-flux measurements of a given light curve. Realizing that the variability is not clearly 
detected (weak intrinsic amplitude or low signal-to-noise ratio) in the multi-wavelength light curves shown in 
Figure \ref{fig:lc}, the uncertainty in F$_{var}$ can be estimated using the expression \citep{Vaughan2003}
\begin{equation}
       \Delta 	F_{var}=\frac{1}{F_{var}} \sqrt{\frac{1}{2N}} \frac{\langle \sigma_{err}^{2}\rangle}{{\langle F \rangle}^2}
\end{equation}   
We have estimated the values of F$_{var}$ for all the energy bands from radio to HE $\gamma$--rays during the period June 15-30, 2014 
(MJD 56823-56838) of Mrk 501 observations. It is to be noted that for \emph{Fermi}-LAT data, F$_{var}$ has been calculated using only 
the detected flux points and excluding the upper limits as shown in the light curve. The values of F$_{var}$ are obtained to be 
negligibly small for OVRO, \emph{Swift}-UVOT, MAXI, \emph{Swift}-BAT and \emph{Fermi}-LAT observations during the above period. 
However, the soft X-ray emission in the energy range 0.3-10 keV observed with \emph{Swift}-XRT is found to be significantly varying with 
the value of F$_{var}$ $=$ 0.23$\pm$0.07 during the VHE flaring activity. The value of F$_{var}$ $=$ 0.11$\pm$0.03 has been estimated for 
VHE $\gamma$--ray emission in the energy range 2--20 TeV during the flaring state of the source observed for $\sim$ 2.1 hours on the night 
of June 23-24, 2014 (MJD 56831.86-56831.94) by the H.E.S.S. telescopes \citep{Chakraborty2015}. It is observed that the estimated value of 
F$_{var}$ for the soft X-ray variability is compatible with the VHE variability observed quasi-simultaneously by H.E.S.S. during the flaring 
activity. However, it can be noted that the estimated value of F$_{var}$ for a given light curve also depends on the size of the time bin. 
A light curve with smaller time bin can give higher value of F$_{var}$, whereas larger time bins can smooth out the flux variations in the 
light curve and yield lower value of F$_{var}$. The values of F$_{var}$ estimated for soft X-rays and TeV $\gamma$--rays support the 
synchrotron and  SSC emission models for the blazars like Mrk 501. 
     
\begin{figure}
\begin{center}
\includegraphics[width=0.80\textwidth,angle=-90]{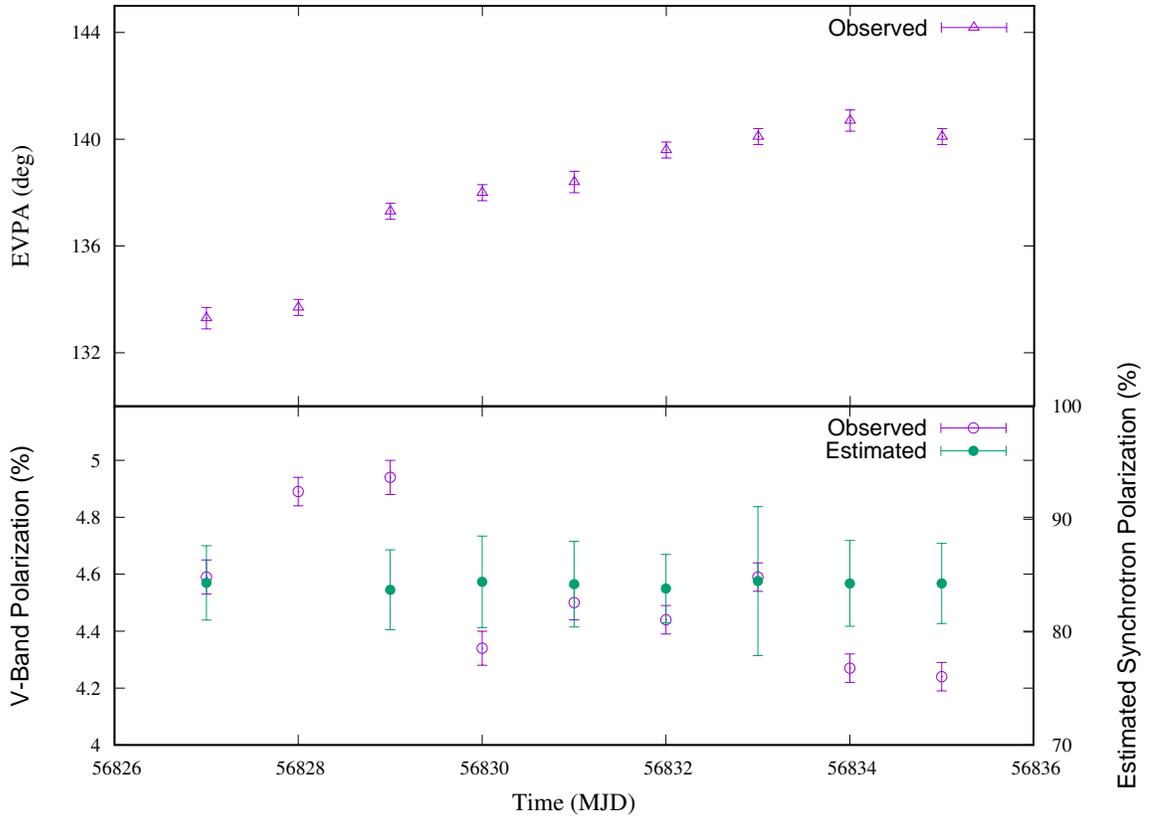}
\caption{Optical linear polarization in V-band (bottom) and electric-vector position angle (top) as a function of time 
         measured at Steward Observatory during the period June 15-30, 2014 (MJD 56823-56838). The short term TeV flaring activity of Mrk 501 is 
        detected by the H.E.S.S. telescope on the night of June 23-24, 2014 (MJD 56831.86-56831.94). The maximum synchrotron polarization estimated from 
        \emph{Swift}-XRT observations during the above period are also shown in the lower panel.}
\label{fig:pol}
\end{center}
\end{figure}

\subsection{Optical Polarization and TeV flare}

Polarization studies of relativistic outflows in extreme astrophysical environments like blazar jets provide important tools to 
probe the geometry of the magnetic field in the emission region and to identify physical processes for the non-thermal emission. 
The observables of polarization of electromagnetic radiation emitted from blazars are degree of polarization ($\Pi$) and electric 
vector position angle (EVPA) or simply polarization angle ($\theta$). The degree of polarization is the fraction of total flux which 
is polarized, whereas EVPA provides information about the orientation of electric field vector in the sky plane. The optical 
polarization also directly indicates the ordering of magnetic field in the jet. The time variation of degree of polarization 
in the optical V-band and associated EVPA measured from Mrk 501 by the \emph{SPOL} telescope during the period 
June 15-30, 2014 (MJD 56823-56838) is shown in Figure \ref{fig:pol}. The degree of polarization in the V-band optical emission of 
the blazar Mrk 501 is observed to be constant with an average value of $\sim$ 4.5$\%$, whereas EVPA changes from $\sim$ 133$^\circ$ to 
a maximum of $\sim$ 141$^\circ$ with a rotation rate of $\sim$ 1$^\circ$ per day during the above period. It is to be noted that 
the degrees of polarization from \emph{SPOL} measurements shown in Figure \ref{fig:pol} are not corrected for thermal emission from 
host galaxy. The variation in EVPA is nearly simultaneous with the flaring activity of the source in the soft X-ray and TeV energy bands. 
The observed possible coincidence of soft X-ray and VHE $\gamma$--ray flares with change in optical polarization angle provides evidence 
for involvement of the synchrotron and SSC processes in the broad-band emission of the source. 
\par
The measurement of synchrotron polarization is the standard approach to investigate the properties of magnetic field and energy distribution of 
emitting electrons in the blazar jet \citep{Westfold1959, Itoh2016}. If the underlying energy distribution of leptons (electrons and positrons) 
emitting synchrotron radiation is described by a power law with spectral index $p$, the maximum degree of linear polarization for a given direction 
of the magnetic field (ordered) is given by \citep{Westfold1959, Rybicki1986, Marscher2014} 
\begin{equation}
	\Pi_{syn}=\frac{p+1}{p+7/3}
\end{equation} 
The spectral distribution of synchrotron photons emitted from the source is also described by a power law with spectral index $\alpha$, 
which  is related to $p$ as 
\begin{equation}
	\alpha=\frac{p-1}{2}
\end{equation}
Equation 3 gives maximum degree of polarization for the synchrotron radiation in an ordered magnetic field. We have estimated the maximum degree of 
polarization ($\Pi_{syn}$) for synchrotron radiation using the values of photon spectral indices ($\alpha$) obtained from the \emph{Swift}-XRT analysis 
(Table \ref{tab:xrt}) in equations 3 \& 4. The values of maximum synchrotron polarization calculated from the soft X-ray observations have been compared 
with the V-band optical polarization measured by \emph{SPOL} at Steward Observatory in Figure \ref{fig:pol} (bottom panel). It is evident from 
Figure \ref{fig:pol} that the maximum synchrotron polarization ($\Pi_{syn}$) during the period of our study has an average value of more than 80$\%$, 
which is much higher than the observed V-band optical polarization with an average value of  $\sim$ 4.5$\%$. The large difference between observed and 
estimated maximum polarization values can be attributed to the various physical effects like tangled and inhomogeneous magnetic field \citep{Gruzinov1999}, 
Faraday rotation and presence of thermal components in the optical emission from the host galaxy \citep{Netzer2013}. {\bf The host galaxies of low redshift 
blazars like Mrk 501 are huge and luminous with elliptical morphology \citep{Hyvonen2007,Nilsson2007}. The previous optical studies of Mrk 501 indicate that the subtraction 
of host galaxy contamination is correlated with the observed magnitudes of the source. A reasonable host galaxy subtraction has been found from the optical 
photometry of the blazar Mrk 501 \citep{Feng2017}. Therefore, the thermal emission from the host galaxy of Mrk 501 has strong impact on the observed polarization 
of the source}. Also, no clear correlation is observed between the V-band optical flux and degree of polarization during the period of TeV flaring activity detected  
by the H.E.S.S. telescopes. This can be attributed to the presence of turbulent magnetic field in the jet emission region. Detailed study of the depolarization effects 
like random magnetic field in the jet emission region and thermal contamination from the host galaxy emission is beyond the scope of this work.  

\begin{figure}
\begin{center}
\includegraphics[width=0.80\textwidth,angle=-90]{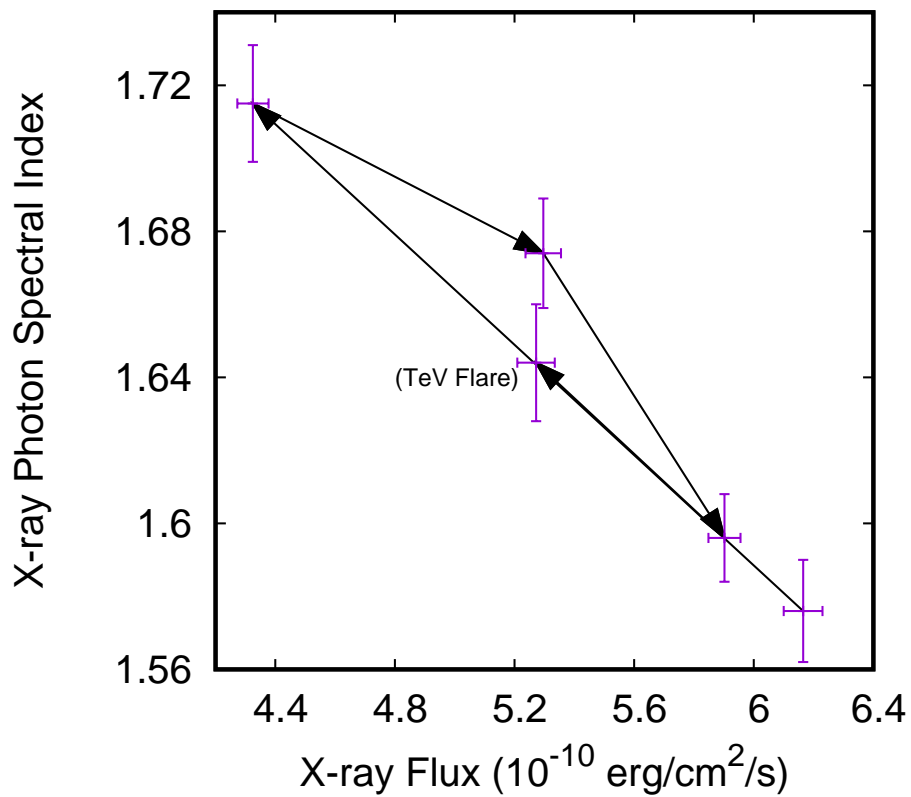}
\caption{Spectral evolution of the soft X-ray emission from Mrk 510 in the energy range 0.3--10 keV measured by \emph{Swift}-XRT during    
        period June 21-24, 2014 (MJD 56829-56832). The arrows indicate time progession of the data points in clock-wise direction during 
        the above period. The data point labelled with TeV flare corresponds to nearest simultaneous XRT observation of the high activity state 
        of Mrk 501 detected by the H.E.S.S. telescopes.}
\label{fig:spec_evol}
\end{center}
\end{figure}
 
\subsection{X-ray Spectral Evolution}
The spectral evolution in the blazar emission helps in investigating the acceleration and cooling features in the jet. This is  acheived by analysing the 
hysteresis patterns of the photon spectral index as function of the source integral flux. The spectral evolution of soft X-ray emission observed with 
\emph{Swift}-XRT in the energy range 0.3-10 keV during the VHE flaring activity of Mrk 501 detected  with H.E.S.S. on June 23, 2014 (MJD 56831.86-56831.94) 
is shown in Figure \ref{fig:spec_evol}. The \emph{Swift}-XRT measurements shown in  Figure \ref{fig:spec_evol} correspond to the observation period during 
June 21-24, 2014 (MJD 56829-56832) including the period of VHE flaring activity. We observe that the soft X-ray photon spectral index is anti-correlated 
with the integral flux in the energy range 0.3-10 keV and indicates the presence of a possible clockwise hysteresis loop structure in the spectral index 
versus flux plane {\bf showing harder-when-brighter behavior}. 

\subsection{Estimation of active region parameter from observables}
The source paremeter like Doppler factor ($\delta_D$) can be estimated from the minimum variability timescale observed in the 
emission of the source. The rapid variability, apparent superluminal motion and high bolometric luminosity observed in blazar 
emission indicate that the broad band non-thermal emission of blazars is produced in a compact active region, relativistically 
moving along the jet oriented towards the observer. The flux variability from radio to VHE $\gamma$--rays at timescales from 
seconds to years is an important observational characteristics of blazars powered by supermassive black holes at the center of 
elliptical host galaxy. Blandford-Znajek mechanism suggests that the jets in blazars are powered by the extraction of rotational 
energy from the supermassive black holes at the center \citep{Blandford1977}. For a maximally rotating supermassive black hole 
of mass $M$, the gravitaional radius is given by 
\begin{equation} 
	R_g=\frac{GM}{c^2}
\end{equation}
where $G$ is the universal gravitational constant and $c$ is the speed of light. An observer can see into the hot dense, highly 
magnetized plasma in the inner parts of the \emph{jet forming region}, whose size will be of the order of $R_g$ \citep{Riordan2016}. 
Assuming that the emitted radiation is dominated by the turbulent plasma of the \emph{jet forming region} close to the event horizon, 
the minimum variability timescale in the comoving frame ($t_{var}$) would be comparable to the event horizon light crossing time 
($t_{EH}$) which is given by 
\begin{equation} 
	t_{EH}=\frac{R_g}{c}
\end{equation}
Relativistic boosting in the blazar emission leads to the shortening of the observed variability timescale ($t_{obs}$) as compared to 
the variability timescale in the comoving frame ($t_{var}$) given by
\begin{equation}
	t_{obs}=\left(\frac{1+z}{\delta_D}\right)~t_{var}
\end{equation}      
where $z$ is the redshift of the source. With the assumption that $t_{var} \approx t_{EH}$, the observed minimum variability timescale gives  
\begin{equation}
	\delta_D=\left(\frac{1+z}{t_{obs}}\right)\frac{GM}{c^3}
\end{equation} 
The host galaxy of blazar Mrk 501 harbours a supermassive black hole of mass $\sim$ 2.2$\times$~10$^9$~$M_\odot$ at its center \citep{Barth2002}. 
The flare of Mrk 501 detected by H.E.S.S. on the night of June 23-24, 2014 (MJD 56831.86-56831.94) at TeV energies shows fast flux variations 
on a timescale of $t_{obs}~\sim$ 6 minutes \citep{Chakraborty2015}. Using these observed quantities in equation 8, we have estimated the value of 
Doppler factor $\delta_D ~\sim$ 31 which is compatible with the value of bulk Lorentz factor for very small jet viewing angle. This implies that 
the hypothesis that the \emph{jet forming region} in the blazar Mrk 501 is close to the event horizon of the supermassive black hole at the center of 
its host galaxy. 
   
\section{Summary} 
The main motivation for this work is to study the major TeV flaring activity state of the high energy synchrotron peaked blazar 
Mrk 501 observed by the FACT and \emph{H.E.S.S.} telescopes on the night of June 23-24, 2014 in the multi-wavelength context from radio to GeV 
$\gamma$--rays. The TeV flux variations observed during this unprecedented flaring activity are characterized by the flux doubling 
timescale of few minutes \citep{Chakraborty2015}. This has provided a unique opportunity to explore the physical process involved in 
the jet emission during the high activity state of the source. We have used multi-wavelength data collected during the period June 
15-30, 2014 on Mrk 501 from the observations with various ground and space-based telescopes covering the electromagnetic spectrum 
from radio to high energy $\gamma$--rays. The important points derived from this study are :
\begin{itemize}
\item Near simultaneous multi-wavelength light curves do not show significant change in the source activity from radio to high energy 
      $\gamma$--ray emission except for the soft X-rays in the energy range 0.3-10 keV detected by \emph{Swift}-XRT during the period 
      June 15-30, 2014. 
 
\item In soft X-rays, the highest activity of the source is detected on June 21, 2014 whereas the VHE flaring activity above 2 TeV is 
      observed on the night of June 23-24, 2014 for a very short duration of $\sim$ 2.1 hours. 

\item Variability analysis of the multi-wavelength light curves using fractional variability amplitude $F_{var}$ also indicates that the 
      source exhibits negligible variations in radio, UV/Optical, X-rays in 10-20 keV and 15-50 keV bands and HE $\gamma$--rays in the 
      energy range 0.1-100 GeV. However, soft X-ray emission of the source in the energy range 0.3-10 keV is observed to vary significantly 
      with the variability amplitude value compatible with that estimated for the TeV light curve during the flaring activity. 
      This can be understood in the frame-work of the synchrotron and SSC model with specific physical condition, wherein, the variation 
      in source emission activity is entirely attributed to the sudden change of the particle density in the emission region. In this case, 
      a quadratic correlation is expected between TeV and soft X-ray flux points.

\item The degree of polarization in optical V-band and electric vector position angle are observed to be uncorrelated. The observed optical 
      polarization has an average value of $\sim$ 4.5$\%$ during the TeV flaring activity whereas electric vector position increases gradually 
      with an average rate of $\sim$ 1$^\circ$ per day during this period. The estimated values of synchrotron polarization emitted by a power 
      law distribution of electrons/positrons in an ordered magnetic field are more than 80$\%$ which are much higher than the observed values of 
      V-band optical polarization. These large differences between the two values of degree of polarization may arise due to presence of different 
      conditions like inhomogeneous and tangled magnetic field and dominance of thermal component in the optical emission from the accretion disk. 

\item The minimum variability timescale of the TeV flux points measured by the H.E.S.S. telescopes during flaring state of the source gives Doppler factor 
      $\delta_D$ $\sim$ 31. This indicates that the jet forming region in the HSP blazar Mrk 501 is near to the event horizon of the supermassive black hole at 
      the center of the host galaxy and radiation is emitted from the turbulent plasma in the jet forming region.

\end{itemize}  

The TeV flaring activity of Mrk 501 in June 2014 has been observed by the FACT and H.E.S.S. telescopes with a flux level comparable to the historical flare of the source detected 
in April 1997 \citep{Djannati1999}. The observed differential energy spectrum in the energy range 330 GeV-13 TeV was described by a log-parabola model with significant 
curvature during the flaring activity detected in April 1997. However, during June 2014 flare, the intrinsic differential energy spectrum of Mrk 501 is represented by 
a simple power law model with no evidence of curvature or cut-off in the energy range 2-20 TeV \citep{Lorentz2016}. Also, more than quadratic correlation between soft 
X-ray and TeV $\gamma$--ray photons was computed during the flaring activity in April 1997 \citep{Katarzynski2005}, iwhereas a linear correlation is violated during 
the TeV flaring activity in June 2014 \citep{Cologna2016}. The differential energy spectra of TeV photons during the flaring activities of Mrk 501 measured on two 
nights in June-July 2005 were also described by log-parabolic function with significant curvature \citep{Albert2007}. The flux variability in the two nights in 2005 
is similar to the one discussed in the present study. A decrease in the optical polarization and change in EVPA by 15$^\circ$ was observed during the strong VHE flaring 
activity of Mrk 501 detected on May 1, 2009 with flux level five times the Crab Nebula flux \citep{Aliu2016}. In the present study, no variation in the optical 
polarization is observed, whereas EVPA changes with a rotation rate of 1$^\circ$ per day during the TeV flaring activity. The optical and HE $\gamma$--ray emissions 
do not show any correlation with the TeV flaring activity considered in the present study. The HE $\gamma$--ray emission indicates low activity state of the source 
during the near simultaneous TeV flaring activity. This can not be explained in the frame-work of simple one SSC model and possibly indicates the presence of two 
emission zones during the flaring activity of the source.

\section*{Acknowledgements} 
We thank the anonymous reviewers for their helpful suggestions and comments to improve the contents of the manuscript. 
We acknowledge the use of public data obtained through \textit{Fermi} Science Support Center (FSSC) provided by NASA. 
This work made use of data supplied by the UK Swift Science Data Centre at the University of Leicester. 
This research has made use of the MAXI data, provided by RIKEN, JAXA and the MAXI team. 
Data from the Steward Observatory spectropolarimetric monitoring project were used. This program is supported by 
Fermi Guest Investigator grants NNX08AW56G, NNX09AU10G, NNX12AO93G, and NNX15AU81G.
Radio data at 15 Ghz is used from OVRO 40 M Telescope and this \emph{Fermi} blazar monitoring program is supported by NASA 
under award NNX08AW31G, and by the NSF under AST-0808050. 
\bibliographystyle{elsarticle-harv}
\bibliography{Mrk-501}

\end{document}